\documentclass[11pt]{article}

\usepackage[margin=1in]{geometry}
\usepackage{amsmath, amssymb}
\usepackage{graphicx}
\usepackage{booktabs}
\usepackage{hyperref}
\usepackage{enumitem}
\usepackage{listings}
\usepackage{xcolor}
\usepackage{tikz}
\usetikzlibrary{shapes.geometric, arrows.meta, positioning}
\usepackage{framed}
\usepackage{caption}
\usepackage[round]{natbib}

\hypersetup{
    colorlinks=true,
    linkcolor=blue!50!black,
    citecolor=blue!50!black,
    urlcolor=blue!50!black
}

\lstdefinestyle{pystyle}{
    language=Python,
    basicstyle=\ttfamily\footnotesize,
    keywordstyle=\color{blue!60!black}\bfseries,
    commentstyle=\color{green!40!black}\itshape,
    stringstyle=\color{red!50!black},
    numbers=left,
    numberstyle=\tiny\color{gray},
    numbersep=6pt,
    frame=single,
    framesep=4pt,
    breaklines=true,
    showstringspaces=false,
    captionpos=b,
    xleftmargin=12pt
}
\definecolor{shadecolor}{rgb}{0.96,0.96,0.92}
\newenvironment{reviewerbox}[1]{%
    \begin{shaded*}\noindent\textbf{Reviewer objection: }\textit{#1}\par\smallskip
}{%
    \end{shaded*}%
}

\title{\textbf{How to Do Statistical Evaluations in ECE/CS Papers:\\
A Practical Playbook for Defensible Results}}

\author{Bhaskar Krishnamachari\\
University of Southern California\\
\texttt{bkrishna@usc.edu}}

\date{}

\begin{document}

\maketitle

\begin{abstract}
Strong experimental papers in electrical and computer engineering and computer science (ECE/CS), especially in systems, networking, and applied machine learning, rest on more than a single impressive number. They rest on a chain of design, measurement, analysis, and validation choices that, taken together, make a result believable. This tutorial is a compact, example-driven guide to that chain for beginning researchers. We organize it as an evaluation \emph{workflow}: claim, hypothesis, unit of analysis, baseline, regime sweep, uncertainty estimate, validation check, and reporting. Within that workflow we cover the classical statistical foundations (descriptive statistics, the central limit theorem, normal- and $t$-based confidence intervals, Student's $t$-test, ANOVA, chi-squared and Pearson correlation, linear regression) alongside the modern, distribution-free techniques (the bootstrap, Wilcoxon and Mann--Whitney tests, Cliff's delta) that are usually preferred for ECE/CS data. We also discuss factorial design, randomization and blocking, multiple-comparison correction, latency-specific pitfalls, simulation verification and validation, equivalence-style claims, and reproducibility. A running example, a comparison of two job-scheduling algorithms on simulated workloads with truncated heavy-tailed job sizes, threads through the tutorial, with Python snippets the reader can paste and adapt. The paper closes with a pre-submission checklist; companion student-facing material (project-type translation tables, an evaluation-plan worksheet, exercises, and a worked ``bad evaluation autopsy'') is collected in a separate workbook released alongside this paper.
\end{abstract}

\section*{How to Use This Guide}
\label{sec:how-to-use}
\addcontentsline{toc}{section}{How to Use This Guide}

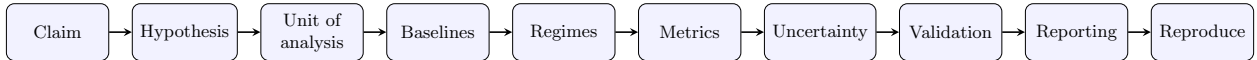
\begin{figure}[h]
    \centering
    \resizebox{\linewidth}{!}{%
    \begin{tikzpicture}[
        every node/.style={font=\footnotesize},
        stage/.style={rectangle, draw, rounded corners, align=center, fill=blue!5, inner sep=4pt, minimum height=10mm, minimum width=18mm},
        arr/.style={-stealth, thick}
    ]
        \node[stage] (claim)  {Claim};
        \node[stage, right=4mm of claim] (hyp)    {Hypothesis};
        \node[stage, right=4mm of hyp]   (unit)   {Unit of\\analysis};
        \node[stage, right=4mm of unit]  (base)   {Baselines};
        \node[stage, right=4mm of base]  (reg)    {Regimes};
        \node[stage, right=4mm of reg]   (met)    {Metrics};
        \node[stage, right=4mm of met]   (unc)    {Uncertainty};
        \node[stage, right=4mm of unc]   (val)    {Validation};
        \node[stage, right=4mm of val]   (rep)    {Reporting};
        \node[stage, right=4mm of rep]   (rpr)    {Reproduce};
        \draw[arr] (claim) -- (hyp);
        \draw[arr] (hyp) -- (unit);
        \draw[arr] (unit) -- (base);
        \draw[arr] (base) -- (reg);
        \draw[arr] (reg) -- (met);
        \draw[arr] (met) -- (unc);
        \draw[arr] (unc) -- (val);
        \draw[arr] (val) -- (rep);
        \draw[arr] (rep) -- (rpr);
    \end{tikzpicture}}
    \caption{The evaluation workflow this tutorial is organized around. Each stage closes a class of reviewer objection; the order matters because choices made early constrain what is provable later (the unit of analysis, in particular, has to be set before the design is locked).}
    \label{fig:workflow}
\end{figure}

The body of this paper walks an evaluation through the chain shown in Figure~\ref{fig:workflow}: from a stated claim, through the design choices that produce evidence, to the analysis and reporting that make it credible. Sections are ordered to match the chain.

This is a 30-page document. Few students will read it in one sitting. Three reading paths cover the common cases.

\paragraph{Undergraduate, first project.} Read \S\ref{sec:weak-vs-strong}--\ref{sec:variability}, \S\ref{sec:effect-size}, \S\ref{sec:latency-pitfalls}, \S\ref{sec:reproducibility}, and Appendix~\ref{sec:checklist}. The goal is to escape single-run, single-number claims.

\paragraph{Master's student writing a systems or ML paper.} Read the whole main text, paying particular attention to the unit-of-analysis discussion (\S\ref{sec:unit-of-analysis}), paired designs, bootstrap CIs, multiple regimes, ablations (\S\ref{sec:ablation}), and reproducibility (\S\ref{sec:reproducibility}).

\paragraph{Beginning PhD student preparing a serious submission.} Read everything, with extra time on randomization and blocking (\S\ref{sec:randomization}), tradeoffs and Pareto fronts (\S\ref{sec:tradeoffs}), simulation verification and validation (\S\ref{sec:sim-validation}), equivalence-style claims (\S\ref{sec:equivalence}), multiple comparisons (\S\ref{sec:multiple-comparisons}), and the rank-test caveats (\S\ref{sec:nonparametric}).

\paragraph{Companion materials.} Three companion artifacts are released alongside this paper at \url{https://github.com/ANRGUSC/statistical-eval-tutorial}:

\begin{itemize}[leftmargin=*, itemsep=1pt, topsep=2pt]
    \item A \emph{student workbook} (\texttt{appendix/student-appendix.pdf}) with a project-type translation table, a fragile-evaluation autopsy, a fill-in evaluation-plan worksheet (with worked example), and exercises mapped to the sections of this paper.
    \item A \emph{reproducible simulator} (\texttt{code/figures.py}) that regenerates every figure and headline number in this paper from a seeded FIFO-vs-SRPT simulation, in a few hundred lines of Python.
    \item A \emph{Claude Code skill} (\texttt{skill/stateval-review/}) that audits a student paper draft against this tutorial's principles and writes a prioritized review document in the working directory; install with \texttt{cp -r skill/stateval-review \textasciitilde/.claude/skills/} and invoke with \texttt{/stateval-review path/to/paper.tex}.
\end{itemize}

\medskip

The single most useful artifact in this document is the following short list. It is the ten-item bar a defensible empirical result has to clear before it leaves your laptop.

\begin{shaded*}
\noindent\textbf{Minimum defensible evaluation.}
\begin{enumerate}[leftmargin=*, itemsep=1pt, topsep=2pt]
    \item State a falsifiable hypothesis.
    \item Choose strong baselines.
    \item Define the unit of analysis (\S\ref{sec:unit-of-analysis}).
    \item Run multiple independent trials/seeds.
    \item Report effect sizes with 95\% confidence intervals.
    \item Use paired comparisons where the design allows.
    \item Test multiple regimes.
    \item Show where the method does \emph{not} help.
    \item Validate the simulator or measurement pipeline.
    \item Release enough information to reproduce the result.
\end{enumerate}
\end{shaded*}

\section{Introduction}
\label{sec:intro}

Evaluation is the part of an experimental paper most likely to be attacked, and most likely to deserve it. A reviewer who is unmoved by your design or your prose will still read the evaluation, and a fragile evaluation will sink an otherwise good paper. Strong evaluations are not about producing a larger improvement; they are about producing a more credible improvement.

A credible result, in our usage, is one that is

\begin{itemize}
    \item statistically sound (variance is reported, tests match the data),
    \item consistent across conditions (the effect does not vanish off the chosen workload),
    \item reproducible (seeds, code, and data permit re-running), and
    \item free of artifacts introduced by the experimental setup (the simulator, the metric, or the pipeline).
\end{itemize}

This tutorial is a practical guide to producing such evaluations in ECE/CS experimental papers. We intend it for student researchers writing their first or second systems, networking, or applied-ML paper, and we have biased it toward the techniques that show up most often in real reviews. It is not a statistics primer. Where a topic has a deep theory, we point to it; we do not develop it. Several earlier works treat overlapping ground at greater length than we can here: \citet{jain1991art} is the canonical book-length reference for systems performance evaluation, \citet{georges2007rigorous} make the case for statistical rigor in benchmark reporting, and \citet{mytkowicz2009wrong} demonstrate how easy it is to produce wrong data through small, plausible-looking measurement choices. For machine-learning evaluations specifically, \citet{demsar2006statistical} treats classifier comparison across many datasets, \citet{dietterich1998approximate} treats statistical tests for paired classifier predictions, and \citet{pineau2021reproducibility} reports on the NeurIPS reproducibility program. We borrow heavily from all of these.

\paragraph{An evaluation workflow, not a stats syllabus.} The body of the tutorial is organized around the chain a credible empirical result has to walk:
\begin{enumerate}[leftmargin=*, itemsep=1pt]
    \item \textbf{Claim.} What is the contribution? Better, faster, more robust, same performance but cheaper, scales better, generalizes?
    \item \textbf{Evidence.} What experiments would convince a skeptical reviewer? Strong baselines, multiple regimes, paired runs, effect sizes, ablations, stress tests.
    \item \textbf{Unit of analysis.} What is one independent observation? Seeds, workloads, datasets, topologies, devices, users, traces.
    \item \textbf{Statistical summary.} Difference in means, in medians, in tail percentiles, in $F_1$, in slopes; with uncertainty.
    \item \textbf{Figure.} What plot makes the claim visible at a glance?
    \item \textbf{Threats.} What can go wrong? Simulator artifact, leakage, cherry-picking, correlated data, bad baseline, too few seeds, metric mismatch.
\end{enumerate}
The sections that follow take this chain in order.

The remainder of the tutorial is organized as follows. Sections~\ref{sec:weak-vs-strong}--\ref{sec:design} set the bar and cover design, baselines, randomization and blocking, and primary-vs-secondary metrics. Sections~\ref{sec:variability}--\ref{sec:hypothesis-testing} address measurement, unit of analysis, confidence intervals (with bootstrap limitations), and hypothesis testing (with caveats on rank tests). Sections~\ref{sec:effect-size}--\ref{sec:regression} cover effect size, equivalence claims, tradeoff/Pareto reporting, and regression. Sections~\ref{sec:multifactor}--\ref{sec:correlated} address multi-factor design, multiple comparisons, out-of-sample validation, ablations, latency pitfalls, reporting, seeds, sample size, outliers, and correlated data. Sections~\ref{sec:sim-validation}--\ref{sec:template} describe simulation verification and validation, machine-learning evaluations, reproducibility, a minimal toolkit, and a template for the evaluation section. Appendices~\ref{sec:checklist}--\ref{sec:glossary} provide a pre-submission checklist and a glossary. Companion student-facing material (project-type translation tables, an evaluation-plan worksheet, exercises, and a worked ``bad evaluation autopsy'') is collected in a separate workbook released alongside this paper.

\subsection*{A running example}

Throughout the tutorial, we use one running example: a comparison of two job-scheduling policies, \textbf{FIFO} (first-in, first-out) and \textbf{SRPT} (shortest-remaining-processing-time), on simulated workloads with Poisson arrivals and heavy-tailed (truncated Pareto) job sizes. We use a Pareto distribution with shape parameter $\alpha = 1.5$ truncated at a maximum job size $x_{\max}$; the truncation is realistic (real systems have finite job sizes, traces have finite length, and simulators impose limits) and avoids a subtle mathematical trap. With untruncated $\alpha = 1.5$, the variance of job size is infinite and the M/G/1 mean response time is theoretically infinite as well; the sample mean then converges very slowly and ``mean latency'' becomes an unstable headline. Truncation keeps the heavy-tail behavior students need to confront while keeping the mean well-defined. The metric of interest is end-to-end \textbf{job completion latency}. This example is small enough to be reproduced in tens of lines of Python, and rich enough to expose every pitfall we discuss: heavy tails (\S\ref{sec:latency-pitfalls}), correlated samples (\S\ref{sec:correlated}), interaction with load (\S\ref{sec:multifactor}), and the gap between simulator behavior and real systems (\S\ref{sec:sim-validation}).

\section{Weak vs.\ Strong Evaluation}
\label{sec:weak-vs-strong}

To set the bar, contrast two ways of reporting the same experiment.

\subsection*{Weak}
\begin{quote}
``SRPT improves mean latency by 60\% over FIFO.''
\end{quote}

This sentence is consistent with at least three different realities: a genuine SRPT win across conditions, a single fortunate seed, or an artifact of an unrealistic arrival rate. The reader cannot tell which.

\subsection*{Strong}
\begin{quote}
``Across 30 random seeds at offered load $\rho \in \{0.3, 0.5, 0.7, 0.9\}$, SRPT reduces mean job completion latency by 63--92\% relative to FIFO under truncated-Pareto job sizes ($\alpha=1.5$); paired Wilcoxon signed-rank tests yield $p < 10^{-8}$ at every load. The relative reduction grows monotonically with offered load, from 63\% at $\rho=0.3$ to 92\% at $\rho=0.9$. Absolute latency reductions are an order of magnitude larger at $\rho = 0.9$ (71 units) than at $\rho = 0.3$ (6 units), so the practical impact is concentrated in the heavy-load regime. The advantage shrinks under exponential job sizes, where SRPT and FIFO are closer in expectation.''
\end{quote}

The strong version is not longer because it is verbose. It is longer because it commits to a number of seeds, a confidence interval method, a paired test, multiple regimes, and the boundary on which the claim depends (the job-size distribution). Each commitment is a place a reviewer might have asked a question, pre-empted.

The rest of this tutorial walks through how to produce evaluations of the second kind without spending more time on statistics than on the underlying contribution.

\section{Experimental Design}
\label{sec:design}

A statistically rigorous analysis cannot rescue a poorly designed experiment. Design comes first.

\paragraph{Hypothesis.} Write the hypothesis the experiment is meant to test before you run it. ``SRPT lowers mean latency relative to FIFO under heavy-tailed workloads'' is testable. ``SRPT is better'' is not, because nothing in the design can falsify it.

\paragraph{Independent and dependent variables.} Identify what you control (offered load, job-size distribution, scheduler) and what you measure (mean latency, p99 latency, throughput). Mixing the two, for example by treating an emergent property as an input, is a common source of confused conclusions.

\paragraph{Baselines.} Compare against the strongest reasonable alternative, not the easiest one. For scheduling, FIFO is the right floor; if a more sophisticated baseline (e.g., processor-sharing) exists in the same regime, include it. A favorable comparison against a strawman is worth almost nothing.

\paragraph{Multiple regimes.} A single working point is not an evaluation; it is an anecdote. For the scheduling example, vary at least the offered load and the job-size distribution. The most important effects in systems work are usually \emph{interactions} between the proposed method and the operating regime (\S\ref{sec:multifactor}).

\paragraph{Repeated trials.} Re-run each configuration with multiple random seeds. A single run reports one realization of a stochastic process; the rest of the tutorial assumes you have repeated trials to work with.

\begin{reviewerbox}{``How do we know this is not cherry-picked to the workload?''}
Pre-commit to the regimes (loads, distributions, sizes) before looking at any results, list them in the paper, and report all of them, including the ones where the proposed method does not win. A short ``where the method does not help'' paragraph is more credible than a uniform sweep of victories.
\end{reviewerbox}

\subsection{Randomization, Blocking, and Order Effects}
\label{sec:randomization}

The seeds-and-multiple-regimes story above assumes the experiment is otherwise clean. On a real measurement testbed it usually is not. Background processes, CPU frequency scaling, thermal throttling, GPU contention, battery state, channel conditions, time of day, and the order in which trials run all introduce \emph{confounders} that can be mistaken for a treatment effect.

Two design moves protect against this. \textbf{Randomization} of the order in which configurations are run breaks any systematic relationship between time-varying confounders and the configuration label: do not run all baseline trials in the morning and all proposed-method trials in the afternoon. Shuffle the trials. \textbf{Blocking} groups trials that share a confounder and varies the treatment within each block. Concretely, if you compare two wireless protocols on ten topologies, do not run protocol A on one set of topologies and protocol B on another. Run both protocols on each topology and each seed, in a randomized within-block order, and analyze the paired differences. The same logic applies to blocking by machine, dataset split, day, channel realization, or workload trace. Blocked-and-paired designs are a major source of statistical power and are what makes the paired tests of \S\ref{sec:hypothesis-testing} appropriate. As a quick checklist for systems-style experiments: pin CPU governor and frequency, disable turbo if measuring stability, fix kernel and library versions, isolate the system under test from background load, and record the environment (temperature, battery, network) alongside the result.

\subsection{Primary and Secondary Metrics}
\label{sec:primary-metric}

Real systems are evaluated on more than one metric: latency and throughput, accuracy and energy, makespan and communication overhead, packet delivery ratio and routing overhead, model quality and training time. A common student failure mode is to report many metrics, then selectively emphasize whichever looks best. Reviewers can detect this and discount the result accordingly.

The discipline is to pick one \textbf{primary metric} tied to the main claim, before running the experiment, and stick with it. Pick a small set of \textbf{secondary metrics} that document the tradeoffs around the primary metric, and report them whether or not they flatter the proposed method. Do not change the primary metric after seeing results, and do not silently drop secondary metrics that came out unfavorably; that is exactly the data-dependent choice the garden of forking paths warns about (\S\ref{sec:multiple-comparisons}). For systems work, it is good practice to include cost or overhead among the secondaries: runtime, memory, communication or message overhead, energy, and implementation complexity. A method that improves the primary metric by 3\% while increasing overhead by an order of magnitude has not unambiguously won, and the evaluation should not hide that.

\section{Measurement and Variability}
\label{sec:variability}

Most ECE/CS experiments are stochastic: random seeds, network jitter, GPU non-determinism, cache state. A single run reports a single sample from a distribution. Treat that distribution as the object of study, rather than the single number.

\paragraph{How many runs.} A practical floor is $n \geq 10$ seeds; $n \geq 30$ is comfortable; for tail metrics ($p_{95}$, $p_{99}$) you generally want substantially more, because each run gives only one estimate of the tail. We treat sample-size selection more carefully in \S\ref{sec:sample-size}.

\subsection{What Is One Data Point? The Unit of Analysis}
\label{sec:unit-of-analysis}

Before any test or interval can be computed, you have to know what counts as one independent observation. Beginning researchers frequently get this wrong, and the resulting confidence intervals can be off by more than an order of magnitude. The pattern looks like this:

\begin{quote}
``We simulated 1{,}000{,}000 packets, so $n = 10^6$. The standard error is tiny.''
\end{quote}

If those packets came from \emph{one} run, with one topology, one mobility trace, one seed, and one channel realization, then there is no sense in which you have a million independent experimental replications. You have one experimental unit, observed a million times. The packet-level samples are correlated within the run, and aggregating them with formulas that assume independence can shrink CIs absurdly. This problem is called \emph{pseudoreplication}: counting non-independent observations as if they were independent.

The fix is to identify the level at which experimental conditions are independently re-rolled, and treat that level as the unit of analysis. For most ECE/CS work the candidates are:

\begin{itemize}
    \item \textbf{Seed.} A new seed redraws the workload, the noise, and (in CRN-style designs) the policy-internal randomness. Seeds are the typical unit for simulation studies.
    \item \textbf{Workload or trace.} If the experiment runs on a held-out collection of real traces (CAIDA flows, MAWI captures, MIMIC-III patients, ImageNet images), each trace is one unit.
    \item \textbf{Topology, machine, or device.} For a wireless evaluation that varies topology or for a benchmark run on multiple machines, each topology/machine is one unit; observations within it are correlated.
    \item \textbf{User or subject.} For systems involving real people (input traces, search queries, evaluation subjects), the user is the unit, even when each user contributes many events.
\end{itemize}

The safe default for a student project is therefore: \emph{summarize each independent run by one number} (e.g., per-seed mean latency, per-seed $p_{99}$, per-seed delivery ratio), \emph{then compute statistics across those run-level summaries}. This collapses millions of correlated raw observations into the much smaller number of genuinely independent experimental replicates, which is the number that should drive standard errors and CIs.

\begin{shaded*}
\noindent\textbf{Wrong vs. right.}
\begin{itemize}[leftmargin=*, itemsep=1pt]
    \item \emph{Wrong:} ``We observed 1{,}000{,}000 packet delays and used a $t$-test with $n = 10^6$.''
    \item \emph{Better:} ``We ran 30 independent seeds. For each seed we computed mean delay, $p_{99}$ delay, and delivery ratio. We report 95\% bootstrap CIs across the 30 seed-level summaries.''
\end{itemize}
\end{shaded*}

When the design is genuinely nested (packets within runs within topologies within scenarios), the analysis must respect that nesting: either pre-aggregate to the highest independent level (the simplest fix), use a clustered or block bootstrap that resamples whole runs/topologies (\S\ref{sec:correlated}), or fit a hierarchical/mixed-effects model that has variance components for each level. For a student paper, pre-aggregation is almost always the right starting move.

This concept reappears throughout the tutorial: it determines what the bootstrap should resample (\S\ref{sec:bootstrap}), what counts as a paired observation (\S\ref{sec:hypothesis-testing}), how ML splits should be drawn (\S\ref{sec:ml}), and what time-correlated data does to standard tests (\S\ref{sec:correlated}).

\subsection{Basic descriptive statistics}
\label{sec:descriptive}

Given $n$ measurements $x_1, x_2, \ldots, x_n$ (for example, mean latency from each of $n$ seeded simulation runs), the standard summary statistics are the sample mean
\begin{equation}
\bar{x} = \frac{1}{n} \sum_{i=1}^{n} x_i,
\end{equation}
the sample variance
\begin{equation}
s^2 = \frac{1}{n-1} \sum_{i=1}^{n} (x_i - \bar{x})^2,
\end{equation}
and the sample standard deviation $s = \sqrt{s^2}$. The denominator $n-1$ (Bessel's correction) makes $s^2$ an unbiased estimator of the underlying variance. The \emph{standard error of the mean} is $\mathrm{SE} = s / \sqrt{n}$; it measures how precisely $\bar{x}$ estimates the true mean, and shrinks as $n$ grows.

For data that is heavy-tailed, asymmetric, or has outliers (latencies often satisfy all three), report robust alternatives alongside the mean: the median (50th percentile), the interquartile range $\mathrm{IQR} = q_{0.75} - q_{0.25}$, and tail percentiles such as $p_{95}$ and $p_{99}$. The mean and SD describe a typical Gaussian-like distribution; the median and IQR describe any distribution.

For each experimental configuration, we recommend reporting at minimum: $n$, the mean (and median, if it differs), a measure of spread (we prefer a confidence interval, see below), and the relevant tail percentiles. Latency is almost never well-summarized by a mean alone (\S\ref{sec:latency-pitfalls}).

\subsection{Confidence intervals}
\label{sec:ci}

A 95\% confidence interval (CI) for a parameter is a random interval that, under repeated experiments, would contain the true parameter value roughly 95\% of the time. CIs are far more informative than standard errors alone because they directly support claims of the form ``$A$ is better than $B$'': if the CI for the difference $\bar{x}_A - \bar{x}_B$ excludes zero, the data support a directional conclusion. We discuss three constructions, ordered from most assumption-laden to most general.

\paragraph{Normal (Wald) CI.} If the data are approximately normal \emph{or} if $n$ is large enough for the central limit theorem to make $\bar{x}$ approximately normal (a common rule of thumb is $n \geq 30$ for moderately skewed data), the textbook CI is
\begin{equation}
\bar{x} \pm z_{1-\alpha/2} \cdot \frac{s}{\sqrt{n}},
\label{eq:wald}
\end{equation}
where $z_{1-\alpha/2}$ is the standard-normal quantile (1.96 for $\alpha = 0.05$). This is the formula taught in introductory statistics courses, and it is the right starting point when its assumptions hold. For latency-like, heavy-tailed data with modest $n$, both assumptions can fail, and the interval may be too narrow (under-coverage).

\paragraph{Student's $t$ CI.} For small $n$ (say, $n < 30$) with approximately normal data, replace the normal quantile with the corresponding quantile from Student's $t$ distribution with $n-1$ degrees of freedom:
\begin{equation}
\bar{x} \pm t_{n-1, 1-\alpha/2} \cdot \frac{s}{\sqrt{n}}.
\label{eq:tci}
\end{equation}
The $t$ quantile is larger than the corresponding normal quantile (e.g., $t_{9, 0.975} = 2.262$ vs.\ $z_{0.975} = 1.96$), giving a wider, more honest interval that accounts for the additional uncertainty in $s$ when $n$ is small. As $n$ grows, $t$ approaches the normal distribution.

\paragraph{Bootstrap CI.}
\label{sec:bootstrap}
For data that is heavy-tailed, ratios, percentiles, or any custom statistic for which Equations~(\ref{eq:wald})--(\ref{eq:tci}) do not apply, the \emph{nonparametric bootstrap} \citep{efron1993introduction} sidesteps distributional assumptions by resampling the observed data. The recipe is short: resample the per-seed measurements with replacement $B$ times (typically $B = 10{,}000$), recompute the statistic of interest on each resample, and take the 2.5th and 97.5th percentiles of the resulting distribution as the 95\% CI.

\begin{lstlisting}[caption={Bootstrap 95\% CI for the mean.},label={lst:bootstrap}]
import numpy as np

def bootstrap_ci(x, stat=np.mean, n_boot=10000, alpha=0.05, rng=None):
    rng = rng or np.random.default_rng(0)
    x = np.asarray(x)
    boots = np.empty(n_boot)
    n = len(x)
    for i in range(n_boot):
        idx = rng.integers(0, n, n)
        boots[i] = stat(x[idx])
    lo, hi = np.quantile(boots, [alpha/2, 1 - alpha/2])
    return stat(x), lo, hi
\end{lstlisting}

The bootstrap is not magic, and it is not assumption-free. It assumes the observed sample is representative of the population of experimental units and that the resampling scheme matches the dependence structure of the data. Several caveats follow from those assumptions and matter in practice:

\begin{itemize}[leftmargin=*, itemsep=1pt]
    \item \textbf{Resample units, not raw events.} Do not bootstrap individual packets, requests, or events when those events are correlated within a run. Resample the unit of analysis (\S\ref{sec:unit-of-analysis}): seeds, runs, traces, or topologies, not the raw observations inside them. Bootstrapping packets pretends each packet is independent and produces CIs that are too narrow.
    \item \textbf{Pair the bootstrap.} For paired comparisons, bootstrap the per-pair differences $d_i$, not the two arms separately.
    \item \textbf{Time series.} For time-correlated data, use the block bootstrap (\S\ref{sec:correlated}); the block length should exceed the dominant autocorrelation scale.
    \item \textbf{Nested data.} For genuinely nested designs (packets in runs in topologies), use a clustered bootstrap that resamples whole top-level units, not individual leaves.
    \item \textbf{Small $n$.} With very small $n$ (say, $n < 8$) the resamples cannot capture the true variability and percentile CIs become optimistic; the $t$-based CI of Equation~(\ref{eq:tci}) is preferable when its assumptions are tenable, and otherwise no method will save the analysis.
    \item \textbf{Extreme tails.} Percentile bootstrap CIs can be inaccurate for very small $n$ or for extreme tail quantities (e.g., $p_{99}$ from a single short run, where the number of \emph{raw} observations per run also matters).
    \item \textbf{Skew.} The bias-corrected and accelerated (BCa) variant adjusts for skew in the bootstrap distribution and is recommended as a refinement when the simple percentile interval looks asymmetric.
\end{itemize}

\paragraph{Which to use.} For ECE/CS experimental data, we recommend the bootstrap as the default because it makes the fewest assumptions and often gives practical, assumption-light intervals for the metrics systems researchers report most often (mean latency, tail percentiles, ratios), provided the resampling unit matches the experimental design (\S\ref{sec:unit-of-analysis}). The normal and $t$-based intervals remain useful as sanity checks, as the right tool when the data is clearly normal (e.g., averages over thousands of independent measurements), and as the formulas every reviewer expects to recognize. In recent versions of SciPy, \texttt{scipy.stats.bootstrap} provides a well-tested implementation including BCa intervals; we show the percentile loop in Listing~\ref{lst:bootstrap} for transparency, but library code is preferable in production.

\begin{figure}[t]
    \centering
    \includegraphics[width=0.7\linewidth]{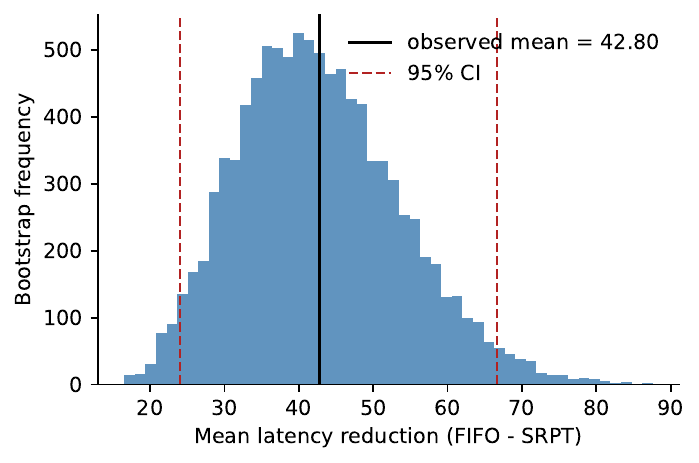}
    \caption{Bootstrap distribution ($B=10{,}000$) of the mean per-seed latency reduction (FIFO $-$ SRPT) at $\rho=0.8$, $n=30$ seeds. The 2.5th and 97.5th percentiles bound the 95\% CI; the bootstrap requires no normality assumption on the per-seed differences.}
    \label{fig:bootstrap}
\end{figure}

\section{Hypothesis Testing}
\label{sec:hypothesis-testing}

Once you have repeated measurements, you can ask whether an observed difference is plausibly due to chance. This section introduces the foundational concepts, then surveys the parametric tests (Student's $t$, ANOVA, chi-squared, Pearson correlation) that beginning researchers are expected to know, then the non-parametric alternatives that are usually preferred for ECE/CS data, then how to check which family applies.

\subsection{Fundamentals}
\label{sec:hyp-fund}

A statistical test compares two hypotheses about the data. The \emph{null hypothesis} $H_0$ is the default position of ``no effect'' (e.g., the two policies have equal mean latency). The \emph{alternative hypothesis} $H_1$ is what we hope to demonstrate (e.g., SRPT has lower mean latency). The test computes a \emph{test statistic} from the data, and the \emph{$p$-value} is the probability, under $H_0$, of obtaining a test statistic at least as extreme as the one observed:
\begin{equation}
p = \Pr(\text{test statistic at least as extreme} \mid H_0).
\end{equation}
A small $p$-value is evidence against $H_0$; a large $p$-value is not evidence \emph{for} $H_0$, only an absence of evidence against it.

The decision rule is to reject $H_0$ if $p < \alpha$, where $\alpha$ is the \emph{significance level}, conventionally 0.05 or 0.01. This rule has two error types: a \emph{Type I error} rejects $H_0$ when it is true (false positive, controlled at rate $\alpha$), and a \emph{Type II error} fails to reject $H_0$ when $H_1$ is true (false negative, controlled by sample size; see \S\ref{sec:sample-size}). The \emph{power} of a test is $1 - \Pr(\text{Type II error})$.

A test is \emph{one-tailed} if $H_1$ specifies a direction (e.g., SRPT lower than FIFO) and \emph{two-tailed} if it does not (e.g., SRPT differs from FIFO). Use two-tailed tests by default, unless you have a strong a-priori reason for a directional hypothesis and committed to it before seeing the data.

\subsection{Parametric tests}
\label{sec:parametric}

Parametric tests assume the data follows a particular distribution (typically normal). When the assumption holds, they are slightly more powerful than their non-parametric counterparts. Student's $t$-test is the most fundamental.

\paragraph{One-sample $t$-test.} Tests whether a sample mean differs from a fixed value $\mu_0$ (e.g., is mean latency below a 100ms SLA?). The test statistic is
\begin{equation}
t = \frac{\bar{x} - \mu_0}{s / \sqrt{n}},
\end{equation}
distributed as Student's $t$ with $n-1$ degrees of freedom under $H_0$.

\paragraph{Paired $t$-test.} Tests whether matched pairs have a mean difference of zero (e.g., FIFO vs.\ SRPT on the same seeds). Compute per-pair differences $d_i = x_i - y_i$ and apply a one-sample $t$-test on the $d_i$ against $\mu_0 = 0$. Pairing dramatically reduces variance when there is per-pair correlation, which is why it is almost always more powerful than treating the two arms as independent.

\paragraph{Two-sample $t$-test.} Tests whether two independent samples have equal means. The classical (Student's) version assumes equal variances:
\begin{equation}
t = \frac{\bar{x}_1 - \bar{x}_2}{s_p \sqrt{1/n_1 + 1/n_2}}, \qquad s_p^2 = \frac{(n_1-1) s_1^2 + (n_2-1) s_2^2}{n_1 + n_2 - 2}.
\end{equation}
Welch's $t$-test relaxes the equal-variance assumption and uses a more complex degrees-of-freedom formula; it should be preferred over the Student's version when in doubt, and is the default in \texttt{scipy.stats.ttest\_ind} when \texttt{equal\_var=False}.

\paragraph{One-way ANOVA.} For three or more independent groups, ANOVA tests whether the group means are all equal. The test statistic is the $F$-ratio of between-group variance to within-group variance, distributed as $F_{k-1, n-k}$ under $H_0$ for $k$ groups and $n$ total observations. ANOVA assumes normality within groups and equal variances across groups; if rejected, follow up with pairwise comparisons (with multiple-comparison correction, \S\ref{sec:multiple-comparisons}) to identify which groups differ.

\paragraph{Chi-squared test.} For categorical data, tests independence between two factors in a contingency table (e.g., does crash rate depend on configuration?). The test statistic compares observed and expected counts under independence:
\begin{equation}
\chi^2 = \sum_{i,j} \frac{(O_{ij} - E_{ij})^2}{E_{ij}}.
\end{equation}
Requires expected counts $E_{ij} \geq 5$ in every cell; for sparse tables, use Fisher's exact test instead.

\paragraph{Pearson correlation.} Measures the strength of a linear relationship between two continuous variables. The sample correlation $r$ ranges from $-1$ to $+1$; $r = 0$ indicates no linear relationship. The test statistic $t = r \sqrt{(n-2)/(1-r^2)}$ has a $t$ distribution with $n-2$ degrees of freedom under $H_0: \rho = 0$. Pearson assumes linearity and approximate bivariate normality; for monotone but non-linear relationships, use Spearman's rank correlation instead.

\subsection{Checking parametric assumptions}
\label{sec:assumptions}

Parametric tests can give misleading $p$-values when their assumptions are violated. Beginning researchers should make the check explicit, rather than assume normality silently.

\paragraph{Q--Q plot.} Plot the sample quantiles against the theoretical quantiles of a normal distribution. Approximately normal data falls on a straight line; heavy tails curve away at the ends; skew shows as systematic curvature. The Q--Q plot is the single most informative normality diagnostic, and the first thing to look at.

\paragraph{Shapiro--Wilk test.} A formal hypothesis test of normality, with $H_0$ that the data is normal. Reject $H_0$ at $\alpha = 0.05$ as evidence of non-normality. Available as \texttt{scipy.stats.shapiro}. With small $n$ the test has low power (will fail to detect non-normality), and with large $n$ it has so much power that even practically unimportant deviations are flagged. Read it together with the Q--Q plot.

\paragraph{Levene's test.} Tests equal variance across groups, useful before applying Student's two-sample $t$-test or ANOVA. Available as \texttt{scipy.stats.levene}. If rejected, use Welch's $t$ or non-parametric alternatives.

If the parametric assumptions hold, use the parametric test. If they do not, use a non-parametric test, the bootstrap, or a transformation (e.g., $\log$ for right-skewed data) that may restore normality.

\subsection{Non-parametric tests}
\label{sec:nonparametric}

Non-parametric tests make minimal distributional assumptions, typically working on the ranks of the data rather than the raw values. They are robust to non-normality, heavy tails, and outliers; the cost is a small loss in power when the parametric assumption actually holds (the asymptotic relative efficiency of Wilcoxon vs.\ paired-$t$ is about 0.95 under normality, and higher under heavy tails). For ECE/CS data, we recommend non-parametric tests as the default.

The defaults that cover most cases are summarized in Table~\ref{tab:tests}, paralleling the parametric tests of \S\ref{sec:parametric}.

\begin{table}[h]
\centering
\footnotesize
\setlength{\tabcolsep}{4pt}
\begin{tabular}{p{0.30\linewidth}p{0.28\linewidth}p{0.32\linewidth}}
\toprule
Scenario & Parametric test & Non-parametric alternative \\
\midrule
One sample vs.\ fixed value & One-sample $t$ & Wilcoxon signed-rank (vs.\ median) \\
Paired, two conditions & Paired $t$ & Wilcoxon signed-rank (on differences) \\
Two independent groups & Two-sample $t$ / Welch's $t$ & Mann--Whitney U \\
Three or more independent groups & One-way ANOVA & Kruskal--Wallis \\
Three or more paired conditions & Repeated-measures ANOVA & Friedman \\
Compare full distributions & (no standard parametric form) & Kolmogorov--Smirnov \\
Linear association & Pearson correlation & Spearman / Kendall rank \\
\bottomrule
\end{tabular}
\caption{Parametric tests and their non-parametric alternatives. For ECE/CS experimental data, we recommend the right column as the default; use the left column when normality is verified or when the data is genuinely Gaussian (e.g., averages of many independent quantities by the central limit theorem).}
\label{tab:tests}
\end{table}

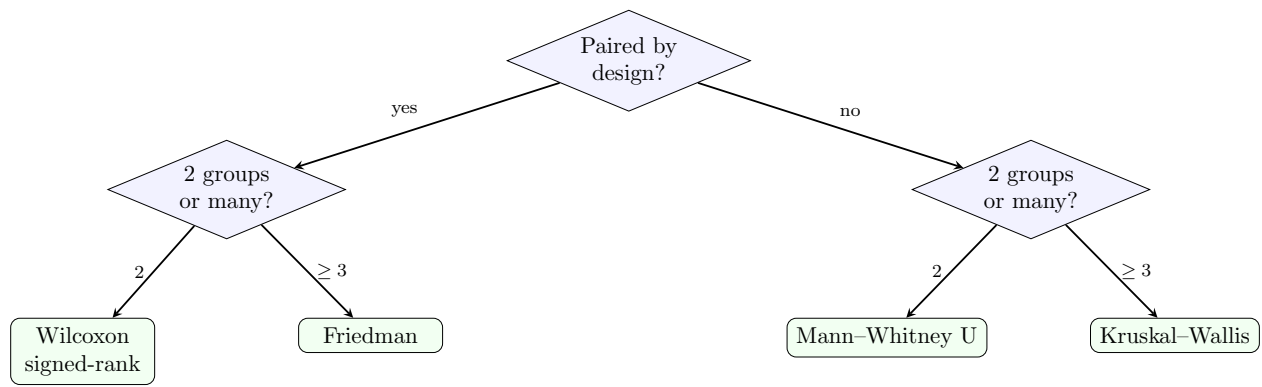
\begin{figure}[t]
    \centering
    \resizebox{\linewidth}{!}{%
    \begin{tikzpicture}[
        node distance=12mm and 28mm,
        every node/.style={font=\small},
        decision/.style={diamond, draw, aspect=2.4, inner sep=2pt, align=center, fill=blue!5},
        action/.style={rectangle, draw, rounded corners, align=center, fill=green!5, inner sep=4pt, minimum width=22mm},
        arrow/.style={-stealth, thick}
    ]
        \node[decision] (paired) {Paired by\\design?};
        \node[decision, below left=of paired, xshift=-15mm] (n2) {2 groups\\or many?};
        \node[decision, below right=of paired, xshift=15mm] (n2u) {2 groups\\or many?};
        \node[action, below=of n2, xshift=-22mm] (wsr) {Wilcoxon\\signed-rank};
        \node[action, below=of n2, xshift=22mm] (fried) {Friedman};
        \node[action, below=of n2u, xshift=-22mm] (mwu) {Mann--Whitney U};
        \node[action, below=of n2u, xshift=22mm] (kw) {Kruskal--Wallis};

        \draw[arrow] (paired) -- node[above left, font=\scriptsize] {yes} (n2);
        \draw[arrow] (paired) -- node[above right, font=\scriptsize] {no} (n2u);
        \draw[arrow] (n2) -- node[left, font=\scriptsize] {2} (wsr);
        \draw[arrow] (n2) -- node[right, font=\scriptsize] {$\geq 3$} (fried);
        \draw[arrow] (n2u) -- node[left, font=\scriptsize] {2} (mwu);
        \draw[arrow] (n2u) -- node[right, font=\scriptsize] {$\geq 3$} (kw);
    \end{tikzpicture}}
    \caption{Default-test selection for ECE/CS data, all non-parametric. Use a $t$-test or ANOVA only after explicitly checking normality (\S\ref{sec:assumptions}) and reporting that you did.}
    \label{fig:flowchart}
\end{figure}

Figure~\ref{fig:flowchart} summarizes the non-parametric defaults as a flowchart for the most common case: comparing two or more groups under uncertainty about distributional shape.

\begin{lstlisting}[caption={Paired Wilcoxon test on per-seed differences (running example).},label={lst:wilcoxon}]
import numpy as np
from scipy.stats import wilcoxon

# fifo, srpt: shape (n_seeds,) -- one mean latency per seed, paired by seed.
diff = fifo - srpt
stat, pval = wilcoxon(fifo, srpt)            # paired
rel = (diff / fifo).mean() * 100             # mean relative reduction
print(f"reduction = {rel:.1f}%, p = {pval:.2e}")
\end{lstlisting}

\paragraph{A more complete decision aid.} The flowchart above selects between two-/multi-group rank tests under uncertainty about distributional shape. Real comparisons need a few more decisions on top of it.

\begin{enumerate}[leftmargin=*, itemsep=1pt]
    \item \emph{Is the same workload/seed/topology used for both methods?} Yes $\to$ paired comparison; no $\to$ independent comparison.
    \item \emph{Two groups or more than two?} Two $\to$ Wilcoxon/paired-$t$ or Mann--Whitney/Welch; more $\to$ Friedman/repeated-measures ANOVA or Kruskal--Wallis/ANOVA.
    \item \emph{Is the claim about the mean, median, a tail percentile, the full distribution, or a probability of an event?} The statistic and CI depend on which one.
    \item \emph{Are observations correlated over time or nested?} Aggregate to run-level summaries, use a block or clustered bootstrap, or fit a mixed-effects model.
    \item \emph{Are you making many comparisons?} Pre-declare the primary comparison, and apply a correction (\S\ref{sec:multiple-comparisons}) to the rest.
\end{enumerate}

\paragraph{Caveats on rank tests.} The convenience of rank tests has a price, and a student who internalizes ``always Wilcoxon'' will eventually overlearn the rule. A few caveats are worth keeping in mind:

\begin{itemize}[leftmargin=*, itemsep=1pt]
    \item Wilcoxon signed-rank assumes the distribution of paired differences is roughly \emph{symmetric}. If symmetry is doubtful and the claim is about the direction of the median, the simpler \textbf{sign test} is safer (less powerful, but only assumes the median).
    \item Mann--Whitney U is not strictly a ``test of medians'' unless the two distributions have similar shape. More generally, it tests whether values from one distribution tend to be larger than values from the other.
    \item Kolmogorov--Smirnov compares full distributions but is less sensitive in the tails and behaves awkwardly with discrete data or many ties.
    \item For paired ML predictions on the same test set, McNemar's test or a paired bootstrap/permutation test on per-example scores is usually more appropriate than treating the two methods' scores as independent \citep{dietterich1998approximate}.
\end{itemize}

\section{Effect Size and Practical Significance}
\label{sec:effect-size}

A small $p$-value tells you the difference is unlikely under the null hypothesis. It does not tell you the difference is large. With enough seeds, any non-zero difference becomes ``significant.'' The corresponding error in reporting is to lead with $p$ and bury the magnitude. Always lead with the effect size, and use the $p$-value to support it.

\paragraph{Estimation first, testing second.} The primary object of reporting is the estimated effect and its uncertainty; the $p$-value is secondary. Instead of asking ``is SRPT significantly better than FIFO?''~the better question is ``how much better is SRPT than FIFO, under which regimes, with what uncertainty, and is the effect practically meaningful?'' Concretely, we recommend the following reporting order for any headline comparison:

\begin{enumerate}[leftmargin=*, itemsep=1pt]
    \item \textbf{Absolute effect}, in natural units: ``mean latency reduced by 70\,ms.''
    \item \textbf{Relative effect}, where it aids interpretation: ``a 62\% reduction.''
    \item \textbf{Confidence interval} on the difference: ``95\% bootstrap CI: 45--95\,ms.''
    \item \textbf{Regime}, where the claim holds: ``at $\rho=0.8$ under truncated-Pareto job sizes.''
    \item \textbf{Test result}, last and shortest: ``paired Wilcoxon $p < 10^{-6}$.''
\end{enumerate}

A reader who can only skim the first three items has the substance of the result; the regime and test add support. A reader who can only see the $p$-value has nothing they can act on.

\paragraph{Cohen's $d$.} The standardized mean difference, useful when comparing across studies with different units:
\begin{equation}
d = \frac{\bar{x}_1 - \bar{x}_2}{s_p},
\end{equation}
where $s_p$ is the pooled standard deviation. Conventional benchmarks (Cohen): $|d| = 0.2$ small, $0.5$ medium, $0.8$ large. Treat the benchmarks as rough guides, not as a substitute for domain judgment.

\paragraph{Cliff's delta.} Non-parametric effect size for two ordinal samples, equal to $\Pr(X_1 > X_2) - \Pr(X_1 < X_2)$. Range $[-1, 1]$; robust to heavy tails and outliers, and the natural companion to Mann--Whitney U.

\paragraph{$\eta^2$ (eta-squared).} For ANOVA, the proportion of total variance explained by the group factor. $\eta^2 = 0.01$ small, $0.06$ medium, $0.14$ large.

\paragraph{Domain-natural effect sizes.} For systems work, a percent improvement in latency, throughput, or energy is usually more interpretable to readers than $d$ or $\eta^2$. Always report the absolute and relative differences, with confidence intervals on the difference. The standardized measures are useful supplements, particularly for cross-study comparison.

The ASA's 2016 statement on $p$-values \citep{wasserstein2016asa} is the canonical reference for the broader point: $p < 0.05$ is a screening criterion, not a conclusion.

\begin{reviewerbox}{``You report $p < 0.001$ but the improvement is 0.3\%.''}
This is the effect-size objection. The fix is to lead with the effect size and its CI, and report the $p$-value in support, rather than the other way around.
\end{reviewerbox}

\section{Tradeoffs and Pareto Fronts}
\label{sec:tradeoffs}

Many ECE/CS papers do not claim a single-metric win. The honest story is a tradeoff: latency vs.\ energy, accuracy vs.\ inference cost, throughput vs.\ fairness, makespan vs.\ communication overhead, packet delivery vs.\ routing overhead, model quality vs.\ training time, robustness vs.\ resource usage. When this is the case, do not pretend one metric dominates. Plot the tradeoff.

A scatter plot with the primary metric on one axis and the cost/overhead metric on the other, with each method's points connected by its parameter sweep (e.g., model size, batch size, scheduling threshold), is usually the most informative view. The relevant comparison is then between the \emph{Pareto fronts} of the methods. A method is \textbf{Pareto-dominated} if some other method is at least as good on every metric and strictly better on at least one; the points on the lower-left frontier of a latency--energy plot, for example, are the ones a designer would actually pick from. Reporting Pareto fronts (rather than a single operating point per method) does two things: it forecloses the objection that the comparison cherry-picked a flattering operating point, and it gives the reader a vocabulary to evaluate the result against their own constraints.

\section{Equivalence and Non-Inferiority Claims}
\label{sec:equivalence}

A surprisingly large number of ECE/CS papers do not actually claim a performance \emph{improvement}. They claim something else:

\begin{itemize}[leftmargin=*, itemsep=1pt]
    \item same accuracy with less energy or compute,
    \item same throughput with fewer messages or lower memory,
    \item same reliability with simpler implementation,
    \item similar performance under more realistic adversarial or failure conditions.
\end{itemize}

Standard significance tests are poorly suited to ``same as baseline'' claims, because the framework is designed to reject a null hypothesis of equality, not to support it. A non-significant $p$-value is \emph{not} evidence that two methods are equivalent; it is consistent with both equivalence and insufficient power.

The right framing is an \textbf{equivalence} or \textbf{non-inferiority} claim with a pre-declared margin. The recipe is short: before running, declare a margin $\Delta$ that captures the largest performance gap you would be willing to call ``the same'' (for example, ``within 2\% of baseline accuracy''). Compute the 95\% CI on the difference, and check whether the entire CI lies inside $[-\Delta, +\Delta]$ (equivalence) or above $-\Delta$ (non-inferiority). If it does, you have evidence for the equivalence/non-inferiority claim at the chosen margin; the CI tells the reader the precision. The margin must be set before seeing the data, and tying it to a substantive criterion (an SLA, an energy budget, a known noise floor) is more defensible than a round number.

Worked example: ``Our lightweight scheduler matches the heavy baseline within $\Delta = 2\%$ mean latency while using 40\% less compute. The 95\% CI on the latency difference is $[-1.1\%, +1.4\%]$, which lies inside the equivalence band; the compute saving has 95\% CI $[36\%, 44\%]$.''

\section{Linear Regression and Correlation}
\label{sec:regression}

When two continuous variables co-vary (e.g., latency vs.\ offered load, energy vs.\ message rate), the basic tools are correlation coefficients and simple linear regression.

\paragraph{Correlation.} Pearson's $r$ measures linear association, Spearman's $\rho$ and Kendall's $\tau$ measure monotone (not necessarily linear) association. Always plot the data first; a scatter plot will reveal whether a linear summary is appropriate or whether the relationship is curved, has clusters, or is dominated by outliers. A high Pearson $r$ on a non-linear relationship is misleading.

\paragraph{Simple linear regression.} Fits $y = \beta_0 + \beta_1 x + \varepsilon$ to data $(x_i, y_i)$ by ordinary least squares (OLS), minimizing $\sum_i (y_i - \beta_0 - \beta_1 x_i)^2$. Report the slope $\hat\beta_1$ with its standard error and a 95\% CI, the intercept, the coefficient of determination $R^2$ (the fraction of variance in $y$ explained by $x$), and a residual plot (residuals vs.\ fitted values; should look like noise around zero with no pattern). Patterns in the residuals indicate that the linear model is mis-specified: a U-shape suggests a missing quadratic term, a fan shape suggests heteroscedasticity, and trends in time suggest autocorrelation. The standard implementation is \texttt{statsmodels.OLS}.

\paragraph{Multiple regression.} Extends to several predictors $y = \beta_0 + \sum_j \beta_j x_j + \varepsilon$. Useful for confounder adjustment and for estimating interactions when factorial design is impractical (\S\ref{sec:multifactor}). Beware of multicollinearity (highly correlated predictors inflate coefficient SEs) and of over-fitting when the number of predictors approaches the number of observations.

\paragraph{Why $R^2$ is not enough.} A high $R^2$ says the model fits the observed data well; it does not say the model is correct, that the relationship is causal, or that the model will generalize. Always report a residual plot alongside $R^2$, and prefer out-of-sample evaluation (\S\ref{sec:out-of-sample}) for predictive claims.

\section{Multi-factor Experiments}
\label{sec:multifactor}

Real systems have more than one knob. Performance depends on offered load, job-size distribution, queue depth, hardware, and the scheduling policy, and these factors \emph{interact}. SRPT may outperform FIFO at high load and tie at low load.

\paragraph{One-factor-at-a-time (OFAT) sweeps.} The most common starting point in practice is to fix sensible defaults for every parameter (load $\rho = 0.7$, job-size shape $\alpha = 1.5$, queue depth $= 1024$, etc.), then sweep one parameter at a time while holding the others fixed. This produces a sequence of one-dimensional plots: latency vs.\ load at default size distribution, latency vs.\ size distribution at default load, and so on. OFAT sweeps are cheap to run, easy to plot, and useful for exposing the effect of each individual parameter. They are the right starting point when you have many parameters and limited time, and they give the reader a quick mental model of where the interesting regimes are.

The danger of OFAT is that it cannot detect \emph{interactions}. If SRPT helps at high load only when sizes are also heavy-tailed, an OFAT sweep over load at light-tailed defaults will miss the effect entirely; an OFAT sweep over size distribution at low-load defaults will too. A method whose advantage requires two factors to align cannot be discovered by one-at-a-time experimentation.

\paragraph{Factorial designs.} A full factorial design with $k$ factors at $L$ levels each requires $L^k$ configurations. For two factors at three levels (e.g., load $\in \{0.3, 0.7, 0.9\}$ and job-size variability $\in \{\text{light}, \text{medium}, \text{heavy}\}$), nine cells is manageable and informative. Plot the resulting $3 \times 3$ grid as an interaction plot; non-parallel lines indicate that the effect of the policy depends on the regime, which is usually the most interesting finding. When the full grid is too expensive, fractional factorial designs sample a structured subset that still permits estimation of main effects and selected interactions; Latin-square-style designs are a related family, useful when two nuisance factors need to be blocked and counterbalanced rather than as a generic fractional factorial method.

\paragraph{Do not average across regimes you care about.} Reporting a single ``mean over all loads'' figure can hide the most important fact about your system, which is \emph{where} it helps. Either present the regimes separately, or use a model (two-way ANOVA, or regression with interaction terms) that estimates the interaction explicitly.

\begin{figure}[t]
    \centering
    \includegraphics[width=0.7\linewidth]{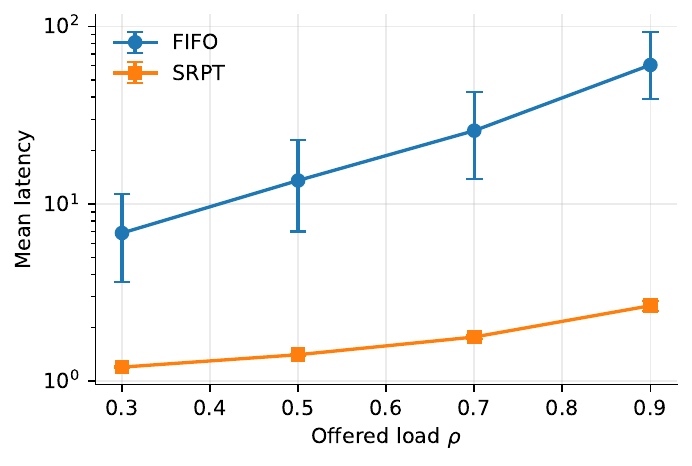}
    \caption{Interaction plot for the running example: mean latency vs.\ offered load for FIFO and SRPT, $n=20$ seeds per cell, 95\% bootstrap CIs. The non-parallel curves on the log-$y$ axis show that the policy effect grows with load. This is the kind of regime-dependent behavior that an averaged single-number summary would hide.}
    \label{fig:interaction}
\end{figure}

\section{Multiple Comparisons}
\label{sec:multiple-comparisons}

Each statistical test you perform has a chance of falsely reporting an effect. Run twenty independent tests at $\alpha = 0.05$ on null data and one will, on average, come up ``significant.'' Papers that report a sweep across many configurations, metrics, and ablations are particularly exposed.

Two corrections cover almost all cases. The Bonferroni correction divides $\alpha$ by the number of tests; it is conservative but trivial to apply. The Benjamini--Hochberg procedure \citep{benjamini1995controlling} controls the false discovery rate (FDR) and is less conservative; it is the right default when you have many tests and are reporting which subset is significant.

The deeper issue is the \emph{garden of forking paths} \citep{gelman2013garden}: when the analyst's choices about which metric, which cutoff, and which subset to report are themselves data-dependent, the effective number of tests is much larger than the number reported. The defenses are pre-registration of the analysis (commit to metrics and cutoffs before running) and reporting every configuration evaluated, including the unfavorable ones.

\section{Out-of-Sample Validation}
\label{sec:out-of-sample}

A method that performs well on the workloads used to design it should also perform well on workloads it has not seen. The cheapest version of this discipline is to hold out: design and tune on one set of workloads, then report final numbers on a different set. Stronger versions test on workloads from a deliberately different regime (different load, different size distribution, different network topology) and on stress conditions chosen to break the method.

In machine-learning evaluations, this discipline is called the train/validation/test split, and the same logic applies in systems work. If every parameter of the proposed method was tuned on the same workload that produces the headline numbers, the headline numbers measure tuning, not generalization.

\section{Ablation Studies}
\label{sec:ablation}

A proposed method is usually a bundle: a new scheduling policy plus an admission control plus a heuristic for tie-breaking; a new neural architecture plus a custom loss plus a data augmentation. The headline number compares the bundle to a baseline. The headline number does not say which component of the bundle is responsible for the gain. An \emph{ablation study} answers that question by removing or replacing one component at a time and measuring how performance changes.

\paragraph{Why ablations matter.} Without ablations, a paper cannot distinguish a genuine new idea from a packaging exercise around an existing one. If 90\% of the gain comes from a well-tuned baseline that the paper happens to bundle with a novel component, the novel component is doing little work. Reviewers know this and ask for ablations precisely to discover it. Conversely, a clean ablation that isolates the contribution of each component is one of the strongest credibility signals an evaluation can provide.

\paragraph{Designs.} The basic ablation patterns are: (i) \emph{leave-one-out}: remove component $C_i$ from the full system and re-run; the gap to the full system measures $C_i$'s contribution. (ii) \emph{add-one-in}: start from the baseline and add components one at a time; this gives a cumulative-gain decomposition. (iii) \emph{component swap}: replace $C_i$ with a simpler or off-the-shelf alternative (e.g., replace a learned scheduler with FIFO, replace a custom loss with cross-entropy) to measure how much the specific design choice matters. Most papers use (i) or (ii); (iii) is the most rigorous when the component has a natural simpler counterpart.

\paragraph{Reporting.} Report ablations in the same conditions and with the same statistical machinery as the main result: same seeds, same regimes, CIs and effect sizes for each removed/replaced component. A common failure mode is to report ablations on a single seed at a single working point; a component that helps on average but hurts in tail conditions will look fine in such a single-cell ablation and unsurprising in deployment. When multiple components are ablated, apply multiple-comparison correction (\S\ref{sec:multiple-comparisons}).

\paragraph{When ablations are not enough.} A clean ablation says ``component $C_i$ contributes $X\%$ on this benchmark.'' It does not say $C_i$ would help on a different benchmark, that the components are independent, or that no further component could be removed without harm. Ablations are evidence about the contribution of components, not proof of design optimality. Pair them with out-of-sample validation (\S\ref{sec:out-of-sample}) for stronger claims.

\section{Latency-Specific Pitfalls}
\label{sec:latency-pitfalls}

Latency data is the most common metric in systems papers and the most consistently mis-reported. Three traps are worth naming.

\begin{figure}[t]
    \centering
    \includegraphics[width=0.7\linewidth]{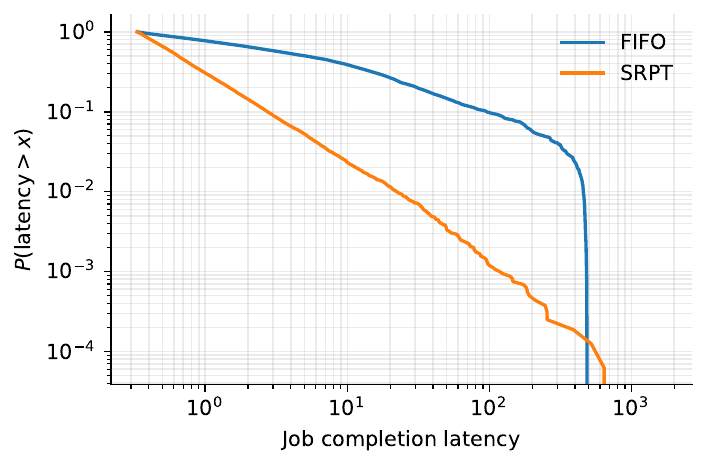}
    \caption{Complementary CDF of job completion latency at $\rho=0.8$, truncated-Pareto-1.5 sizes, log--log axes, $n = 30$ seeds. Plotting the full distribution exposes tail behavior that the mean and median both hide.}
    \label{fig:cdf}
\end{figure}

\paragraph{The mean is not the right summary.} Latency distributions are typically heavy-tailed. The mean is dominated by rare large values; the median ignores them; the user experience usually depends on the tail. Report at least the median and one or two tail percentiles ($p_{95}$, $p_{99}$), and plot the CDF alongside the summary statistics. Figure~\ref{fig:cdf} shows the latency complementary CDF for the two policies on log--log axes; the relevant differences are visible across orders of magnitude in a way that no bar chart of means could convey.

\paragraph{Coordinated omission.} Many load generators measure the time from request \emph{dispatch} to response, but pause dispatch when the system stalls. The result is a latency distribution that looks artificially good under overload, because the worst-case requests were never sent. The fix, due to \citet{tene2015latency}, is to measure latency relative to the request's \emph{intended} dispatch time, not the actual one. Any closed-loop benchmark that does not correct for coordinated omission should be regarded with suspicion.

\paragraph{Warmup and JIT effects.} Cold caches, JIT compilation, lazy connection setup, and resource ramp-up all distort early measurements. Discard or separately analyze a warmup window. Document the warmup duration and how it was chosen.

\section{Reporting and Figure Design}
\label{sec:reporting}

Good figures are evaluations in compressed form; bad figures hide them.

\begin{figure}[t]
    \centering
    \includegraphics[width=0.95\linewidth]{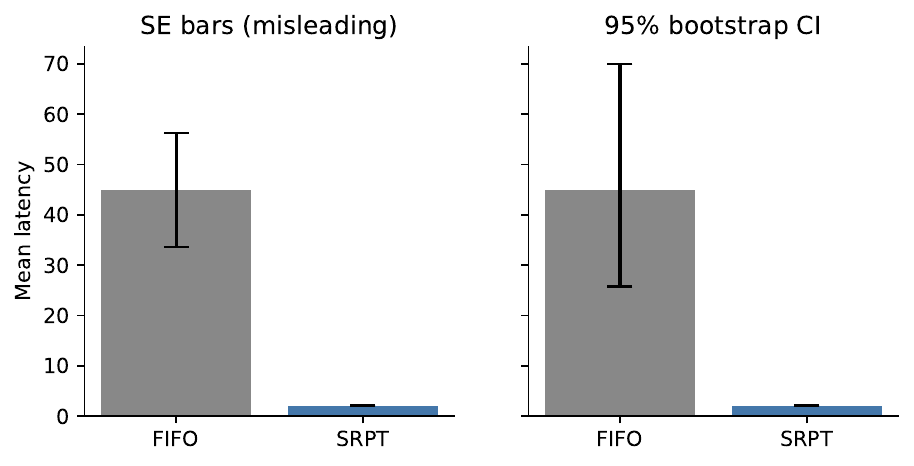}
    \caption{Same 30-seed FIFO vs.\ SRPT comparison plotted with standard-error bars (left) and 95\% bootstrap CIs (right). The SE bars are roughly $\sqrt{n}$ times smaller than the SD and visually compress the variability; the CI bars correctly support the comparative claim.}
    \label{fig:errorbars}
\end{figure}

\paragraph{Default to 95\% CIs, not standard errors or standard deviations.} Beginners frequently confuse four related quantities: the \textbf{standard deviation} (SD) describes the variability of observations or runs; the \textbf{standard error} (SE) of the mean describes how precisely the sample mean estimates the true mean and equals $s/\sqrt{n}$; the \textbf{confidence interval} is the uncertainty interval for a parameter and is roughly $\bar{x}\pm 2\,\mathrm{SE}$ for a mean under approximate normality; and the \textbf{CI of the difference} is what you usually actually want. Standard error bars look small and tempt the reader into reading them as significance bars; they are not. Confidence intervals support the visual claim that ``these conditions differ'' directly. Figure~\ref{fig:errorbars} shows the same data plotted with both kinds of bars; the SE version invites the eye to call the means ``the same,'' which is not what the data say.

\paragraph{When the claim is comparative, plot the CI of the difference.} If the scientific claim is ``A improves over B,'' the cleanest interval is the CI on the paired difference $\bar{x}_A - \bar{x}_B$, not separate error bars on $\bar{x}_A$ and $\bar{x}_B$. Two facts make this important. First, non-overlap of separate 95\% CIs is not exactly equivalent to a significant difference at the 5\% level: two intervals can overlap visibly while the difference is significant, and (less commonly) two intervals can fail to overlap while a paired test does not declare the difference significant. Second, when the experiment is paired (same seeds, same workloads), the CI of the paired difference is much narrower than the CI computed from the two separate arms, because the per-pair correlation cancels out exogenous noise. Reporting only separate bars throws this advantage away. The bootstrap recipe is the same as Listing~\ref{lst:bootstrap}, applied to the per-seed differences $d_i = x_{A,i} - x_{B,i}$.

\paragraph{Use log axes for heavy-tailed quantities.} Plotting latency on a linear axis collapses the body of the distribution and exaggerates the tail. A log $y$-axis (or a complementary CDF on a log $x$-axis) shows the structure.

\paragraph{Show the points when you can.} For $n \leq 20$, a strip plot or jittered scatter alongside the bar reveals what summary statistics hide: bimodal distributions, outliers, runs that crashed.

\begin{figure}[t]
    \centering
    \includegraphics[width=0.95\linewidth]{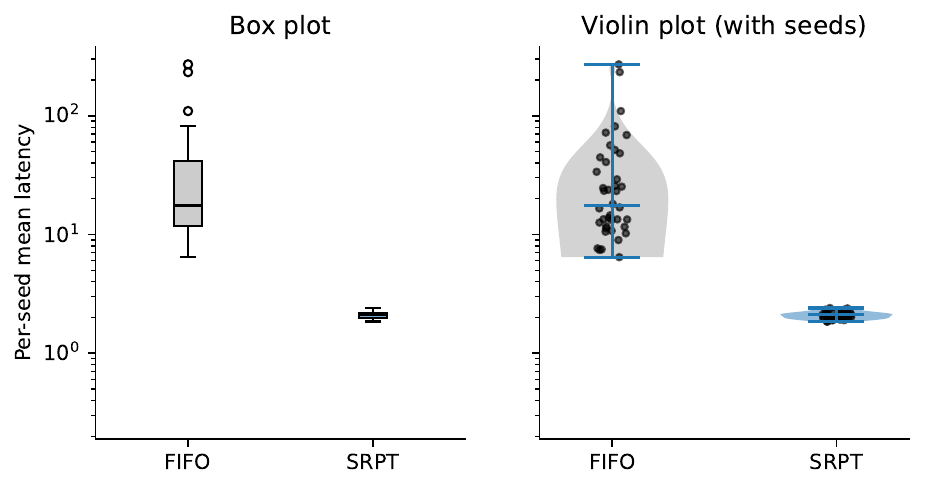}
    \caption{Per-seed mean latency for FIFO and SRPT at $\rho = 0.8$, $n = 40$ seeds, log-$y$ axis. The box plot (left) shows the median, IQR, whiskers, and outliers. The violin plot (right) adds the full kernel-density estimate and overlays the individual seed values, exposing the spread and shape of the distribution that the box hides.}
    \label{fig:boxviolin}
\end{figure}

\paragraph{Box plots and violin plots.} When comparing distributions across a small number of conditions, prefer a box plot or a violin plot over a bar chart. A box plot shows the median, the IQR (the box), the typical spread (the whiskers), and outliers as individual points; it is compact and well-understood, and works well when distributions are roughly unimodal. A violin plot shows a kernel-density estimate of the full distribution shape mirrored about the axis, and is the right choice when the distribution may be bimodal, multimodal, or otherwise non-Gaussian (a long, thin violin is heavy-tailed; two bumps indicate two modes). For $n \leq 40$ or so, overlay the individual seed values as jittered points on the violin to give the reader a direct view of the data. Figure~\ref{fig:boxviolin} contrasts the two on the running example.

\paragraph{Make captions self-contained.} A reader skimming figures should be able to understand each one without returning to the body. Caption: ``Median end-to-end latency vs.\ offered load, $n=30$ seeds, error bars show 95\% bootstrap CI.''

\section{Random Seeds and Determinism}
\label{sec:seeds}

A random seed fixes the sequence of pseudo-random numbers a program consumes. It does not, on its own, fix the result of the program. CUDA kernels, parallel scheduling, floating-point reductions in non-deterministic order, and dependence on wall-clock time all introduce non-determinism that seeding will not remove.

For systems experiments, the practical guidance is: seed every random source you can (workload generator, network jitter model, scheduler tie-breaks), record the seed in the output, and run multiple seeds. For ML experiments on GPUs, full bitwise determinism is often impractical; the alternative is to budget for variance across seeds rather than pretend it does not exist.

\section{Sample Size and Power}
\label{sec:sample-size}

``How many seeds did you run, and why that many?'' is one of the most common reviewer questions. Answer it before being asked.

A statistical \emph{power} analysis estimates the sample size needed to detect an effect of a given size with a given probability. The minimum useful version, for a paired comparison, is: run a small pilot (say, 10 seeds), estimate the per-seed standard deviation $s$ of the difference and the smallest improvement $\Delta$ you would care about, and then use
\[
n \approx \left(\frac{z_{1-\alpha/2} + z_{1-\beta}}{\Delta / s}\right)^2
\]
to estimate the seeds needed for power $1-\beta$ at significance $\alpha$. For $\alpha = 0.05$, $\beta = 0.2$, this is $n \approx 8 / (\Delta/s)^2$. If your effect-to-noise ratio is small, you need many seeds; if it is large, even ten will do.

For tail metrics ($p_{99}$), the relevant quantity is the number of \emph{tail samples} per run, not the number of runs. A run with too few requests gives an unstable $p_{99}$ no matter how many times it is repeated.

\section{Outliers, Failed Runs, and Exclusion Rules}
\label{sec:outliers}

Real experiments produce ``weird'' runs: a run that crashed, a run with an order-of-magnitude latency spike, a seed where the GPU thermally throttled, a trace where the radio briefly disconnected. Beginning researchers often quietly drop these from the analysis. Reviewers, correctly, treat silent exclusion as a credibility threat. A few rules avoid the problem:

\begin{itemize}[leftmargin=*, itemsep=1pt]
    \item \textbf{Define exclusion criteria before running.} A run is excluded if X occurs (the simulator times out, the device crashes, the network disconnects). Write the rule down before you see the results.
    \item \textbf{Do not exclude on the outcome variable.} Excluding a run because the headline metric came out unfavorable is data-dependent selection and violates the analysis assumptions.
    \item \textbf{Report counts.} ``We ran 30 seeds; 2 were excluded due to simulator non-convergence; reported $n = 28$.'' The reader can decide whether the exclusion is suspicious.
    \item \textbf{Show with and without.} If the exclusions move the headline number materially, present both versions. Robust conclusions survive the comparison.
    \item \textbf{Crashes are often results.} A method whose 5\% crash rate is hidden by silent exclusion is much worse than the headline says. The crash rate itself is a metric.
    \item \textbf{Tail values may be the point.} For latency, rare large values are not noise; they are the phenomenon of interest. Trimming the tail to ``stabilize the mean'' redefines the metric.
\end{itemize}

\begin{figure}[t]
    \centering
    \includegraphics[width=0.7\linewidth]{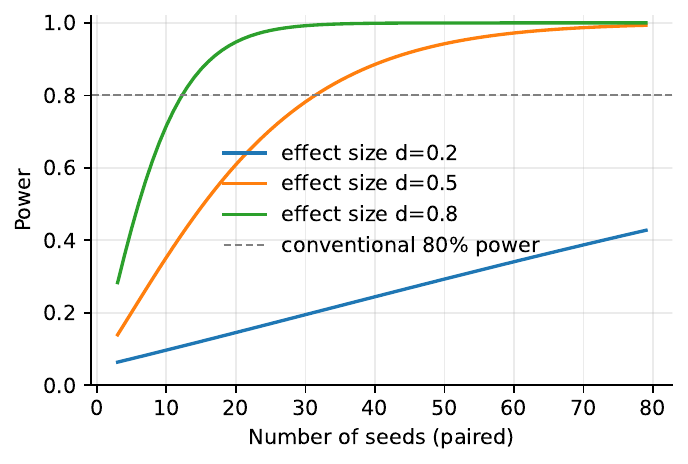}
    \caption{Power vs.\ number of paired seeds, $\alpha=0.05$ two-sided, for three effect sizes (Cohen's $d$). Detecting a small effect ($d=0.2$) reliably needs nearly 80 seeds; a medium effect needs about 30; a large effect is detected with under 15.}
    \label{fig:power}
\end{figure}

\section{Time-Correlated Data}
\label{sec:correlated}

Standard tests and bootstrap CIs assume independent samples. For time-series data (packet inter-arrival times, queue occupancy, throughput sampled every second), adjacent samples are not independent, and naive analysis underestimates variability, sometimes badly.

Two fixes cover most of what student researchers need. The first is \textbf{aggregation}: summarize each run by a single statistic per experiment (mean latency, $p_{99}$, throughput) and treat those summaries as the unit of analysis. This works well when ``one run'' is naturally one independent draw. The second is the \textbf{block bootstrap}: rather than resampling individual time-points, resample contiguous blocks long enough to preserve the autocorrelation. The block length should be chosen larger than the dominant autocorrelation timescale; doubling the block length and checking that the CI is stable is a reasonable diagnostic.

\section{Simulation Verification and Validation}
\label{sec:sim-validation}

Most systems and networking papers rely on simulators. Simulators are convenient and almost always wrong in some way that matters; the question is whether the way they are wrong undermines the claim. The standard discipline distinguishes two related but distinct questions, both of which need answers before any headline result is trusted.

\paragraph{Verification: did we implement the model correctly?} This is an internal-consistency question. Even a perfect model is useless if the code does not implement it. Verification checks include: deterministic toy cases with closed-form answers, limiting cases (zero load, infinite capacity), unit tests for the core data structures, and conservation laws (packets sent = received + dropped, total work done = total work submitted) that must hold to machine precision. These are cheap to write, embarrassing to fail, and frequently fail.

\paragraph{Validation: is the model appropriate for the real system or claim?} This is an external-correspondence question. A correctly implemented model can still be the wrong model: idealized channel models, missing failure modes, unrealistic workloads, or absent hardware effects can each break a claim that depends on them. Validation checks include comparison to analytical results (M/M/1 queue lengths, TCP throughput under known loss), reproduction of published baselines, comparison to measurements on real hardware, agreement with expected scaling laws, and a written \emph{scope statement} that lists what the simulator does \emph{not} model. The scope statement is a load-bearing piece of documentation: it tells the reader (and reviewers) which claims the simulator can support and which it cannot.

A few specific checks cover most of what students need.

\paragraph{Sanity checks (verification).} The simulator should produce trivial answers to trivial questions. Zero offered load should produce zero queueing delay. Infinite link capacity should produce zero drops. Idle workers should not produce non-zero CPU time.

\paragraph{Baseline reproduction (validation).} Run the simulator against a published or analytically known result, such as M/M/1 mean queue length under Poisson arrivals or TCP throughput under a known loss rate, and verify it matches within reasonable tolerance. If it does not, the simulator has a bug, or you are misunderstanding the model.

\paragraph{Input validation.} Are the input distributions, mobility models, channel models, and workload traces realistic for the regime you claim to evaluate? A heavy-tailed claim under exponentially distributed inputs is not what it looks like.

\paragraph{Sensitivity analysis.} Vary each parameter and confirm the response is qualitatively right. Doubling the arrival rate should increase queue length; halving the bandwidth should reduce throughput; turning off the proposed mechanism should give back the baseline. Surprising sensitivities are usually bugs. A stronger version of the same test asks whether the \emph{conclusion} (not just the response surface) survives plausible parameter changes; if a 20\% perturbation of an input distribution flips the result, the headline is fragile.

\paragraph{Internal consistency.} Conservation laws should hold to machine precision. If they do not, the simulator is dropping work somewhere it should not be.

\paragraph{Variance reduction.} Common random numbers (CRN) is an under-used technique in systems papers: when comparing two policies, drive both with the same arrival sequence and the same job-size sequence, varying only the policy. The per-seed difference then has far lower variance than the difference of independent runs. CRN is nearly free in implementation cost, and it directly enables the paired tests of \S\ref{sec:hypothesis-testing}. Implemented carelessly, CRN can desynchronize the workload between policies; Listing~\ref{lst:crn} shows the safe pattern.

\begin{lstlisting}[caption={Common random numbers: generate the workload \emph{once} per seed and replay it under each policy. Seeding two policies with the same RNG is not enough, because the policies may consume random numbers in different orders.},label={lst:crn}]
import numpy as np

def make_workload(seed, rho, n_jobs=5000, alpha=1.5, x_max=1e4):
    # Exogenous randomness: arrivals and (truncated-Pareto) sizes.
    rng = np.random.default_rng(seed)
    arrivals = generate_arrivals(rng, rho, n_jobs)
    sizes    = generate_job_sizes(rng, n_jobs, alpha=alpha, x_max=x_max)
    return arrivals, sizes

def run_pair(seed, rho, n_jobs=5000):
    workload = make_workload(seed, rho, n_jobs)
    # Policy-internal randomness (e.g., tie-breaks) gets its own stream
    # so that one policy's coin flips do not perturb the other's workload.
    fifo = simulate("FIFO", workload, policy_seed=seed + 10_000)
    srpt = simulate("SRPT", workload, policy_seed=seed + 20_000)
    return fifo.completion_times.mean(), srpt.completion_times.mean()

pairs = [run_pair(s, 0.8) for s in range(30)]   # paired by seed
fifo, srpt = map(np.array, zip(*pairs))
\end{lstlisting}

The lesson is that CRN means the \emph{exogenous} randomness (arrivals, job sizes, channel trace, mobility trace, topology) is shared across policies, while each policy's \emph{internal} randomness (tie-breaks, randomized exploration) gets its own independent stream. A single shared RNG handed to both policies will desynchronize the workload as soon as the policies consume random numbers at different rates.

\section{A Note on Machine-Learning Evaluations}
\label{sec:ml}

Systems work increasingly trains or deploys models, and the same playbook applies, with three additional disciplines specific to ML.

\paragraph{Data leakage.} Train, validation, and test data must be separated cleanly. Data leakage, when information from the test set influences the model (through pre-processing computed on the full data, label leakage from features that are themselves derived from the label, temporal leakage when future data informs predictions about past data, or grouped data split across folds), is the single most common cause of inflated reported numbers. Always split before any preprocessing, and verify the split with at least one sanity check (e.g., train on one half of dates, test on the other half).

\paragraph{Splits respect the unit of analysis.} The split also has to respect the structure of the data, not just its size. The unit-of-analysis lesson of \S\ref{sec:unit-of-analysis} reappears here. If rows are correlated by user, device, patient, scenario, or time window, a random row-level split puts correlated examples on both sides of the train/test boundary and inflates test-set performance. The fix is a \emph{grouped split} (split by user/device/scenario, not by row) for clustered data, and a \emph{temporal split} (train on past, test on future) for forecasting and time-series anomaly detection. For hyperparameter tuning, use \emph{nested cross-validation}: an outer split for honest test estimates and an inner split for tuning. Threshold-dependent metrics ($F_1$, accuracy, $p_{99}$ recall at fixed false-positive rate) require the threshold to be picked on validation data, not test data. Confidence intervals on ML metrics are then computed by paired bootstrap over test \emph{examples} (or test \emph{groups}, when the data is clustered), comparing classifiers on the same examples; this is the ML analogue of the paired tests of \S\ref{sec:hypothesis-testing}.

\paragraph{Choose metrics that match the task.} The evaluation metric should match the task and the cost structure of errors. The right metric is more important than the right test, and beginning researchers consistently underweight this point. The defaults by task family are as follows.

\textit{Classification.} For balanced binary classification, report accuracy, precision, recall, and the $F_1$ score (the harmonic mean of precision and recall). For imbalanced data, accuracy alone is misleading and should be replaced or supplemented with the precision-recall (P--R) curve and the average precision (AP), which is the area under the P--R curve. The receiver operating characteristic (ROC) curve and the area under it (AUROC) summarize the trade-off between true positive rate and false positive rate; AUROC is widely reported but can flatter classifiers on highly imbalanced data, where the P--R view is more honest. For multi-class problems, report per-class metrics in addition to macro- and micro-averaged $F_1$. When comparing classifiers across many datasets, \citet{demsar2006statistical} recommends Friedman tests with post-hoc Nemenyi or Bonferroni--Dunn comparisons, rather than per-dataset $t$-tests. For tasks with calibrated probability outputs, also report a calibration curve and a calibration metric (Brier score, expected calibration error).

\textit{Regression and continuous prediction.} Report mean squared error (MSE) or root MSE (RMSE) when large errors are disproportionately costly, mean absolute error (MAE) when they are not, and the mean absolute percentage error (MAPE) when scale-free comparison across targets is needed. The coefficient of determination $R^2$ is useful as a unitless summary; pair it with a residual plot, as in \S\ref{sec:regression}. For probabilistic forecasts, report a proper scoring rule such as the log score or the continuous ranked probability score (CRPS).

\textit{Ranking and information retrieval.} Report normalized discounted cumulative gain (NDCG) at $k$, mean average precision (MAP), and mean reciprocal rank (MRR), choosing among them based on whether graded relevance, binary relevance, or position of the first relevant item is the quantity of interest.

\textit{Language models.} Report perplexity (the exponential of the per-token cross-entropy) on a held-out corpus for autoregressive language modeling. For machine translation, summarization, and other generation tasks, report task-specific scores such as BLEU, ROUGE, METEOR, BERTScore, and chrF, plus a model-graded or human-evaluation pass for tasks where automatic metrics are known to correlate poorly with quality. For factual tasks, report exact-match and $F_1$ against a reference, and consider a contamination check (looking for the test items in the training data). Contamination is a particular concern for evaluations of foundation models trained on web-scale corpora that may already include the benchmark; report the contamination check and, where automatic metrics are known to correlate poorly with quality, supplement with model-graded or human evaluation.

\textit{Detection and segmentation.} Report intersection-over-union (IoU), mean average precision at IoU thresholds (mAP\@0.5, mAP\@[.5:.95]), and per-class AP.

\textit{Reinforcement learning.} Report mean episodic return with variance across seeds (this is where the seed-variance problem of \citet{henderson2018deep,agarwal2021precipice} is most acute), interquartile mean (IQM) as advocated by \citet{agarwal2021precipice}, and stratified bootstrap CIs across seeds.

\paragraph{Variance across seeds.} Recent results on ML evaluation reproducibility, particularly in reinforcement learning, have demonstrated that variance across random seeds is large enough to invalidate many published comparisons \citep{henderson2018deep,agarwal2021precipice}. The defenses are the same as in this tutorial: more seeds, paired non-parametric tests, effect sizes with CIs, and reporting where the method does not work.

\section{Reproducibility}
\label{sec:reproducibility}

A paper is reproducible to the extent that an independent reader, with reasonable effort, can re-run it and get the same numbers. The minimum bar:

\begin{itemize}
    \item state the number of runs, the seeds, and the hardware;
    \item state the parameter-tuning procedure (and apply it to baselines as well as the proposed method);
    \item include all ablations referenced in the text;
    \item release the code and the workload-generation scripts.
\end{itemize}

A reproducibility appendix or artifact submission costs little after the fact and substantially raises the credibility of the result. The NeurIPS reproducibility program report \citep{pineau2021reproducibility} is a useful checklist source for ML work.

\section{A Minimal Statistical Toolkit}
\label{sec:toolkit}

The methods in this tutorial reduce to a small kit. If you can apply these competently, you can defend most experimental papers.

\paragraph{Classical foundations (recognize and use when applicable).}
\begin{itemize}
    \item Sample mean, variance with $n-1$ denominator, standard deviation, standard error (\S\ref{sec:descriptive}).
    \item Normal-based and Student's $t$ confidence intervals; central limit theorem (\S\ref{sec:ci}).
    \item One-sample, paired, and two-sample (Welch's) $t$-tests; one-way ANOVA; chi-squared; Pearson correlation (\S\ref{sec:parametric}).
    \item Q--Q plots and Shapiro--Wilk for checking normality before applying parametric tests (\S\ref{sec:assumptions}).
    \item Simple and multiple linear regression with $R^2$ and residual diagnostics (\S\ref{sec:regression}).
\end{itemize}

\paragraph{Defaults for ECE/CS data.}
\begin{itemize}
    \item Bootstrap confidence intervals (\S\ref{sec:bootstrap}).
    \item Wilcoxon signed-rank (paired) and Mann--Whitney U (unpaired); Kruskal--Wallis for $\geq 3$ groups (\S\ref{sec:nonparametric}).
    \item Effect sizes: relative \% improvement, Cohen's $d$, Cliff's delta, with CIs (\S\ref{sec:effect-size}).
    \item Factorial sweep over at least one regime variable (\S\ref{sec:multifactor}).
    \item Multiple-comparison correction (Bonferroni, Benjamini--Hochberg) when reporting many tests (\S\ref{sec:multiple-comparisons}).
    \item Block bootstrap for time-series data (\S\ref{sec:correlated}).
    \item Simulation sanity checks and baseline reproduction (\S\ref{sec:sim-validation}).
\end{itemize}

\section{Evaluation Section Template}
\label{sec:template}

A defensible evaluation section follows roughly this structure. Treat it as a checklist for self-review before submission.

\begin{enumerate}
    \item \textbf{Experimental setup.} System under test, workloads, parameters, seeds, hardware.
    \item \textbf{Metrics.} Mean and tail; CIs; how variability is computed.
    \item \textbf{Main results.} Comparisons with error bars; effect sizes with CIs.
    \item \textbf{Statistical validation.} Tests applied, $p$-values reported alongside effect sizes.
    \item \textbf{Robustness.} Multiple regimes; where the method does and does not help.
    \item \textbf{Out-of-sample tests.} Held-out conditions; stress tests.
    \item \textbf{Simulation validation.} Sanity checks; baseline reproduction; sensitivity.
    \item \textbf{Ablation study.} Each component's contribution isolated.
    \item \textbf{Summary.} Precise, bounded conclusions.
\end{enumerate}

\section{Conclusion}

The credibility of an evaluation rests on the evidence behind it, rather than on the magnitude of the headline number. Each technique in this tutorial closes a specific class of reviewer objection: variance reporting closes ``maybe you got lucky''; paired tests close ``maybe it is noise''; multiple regimes close ``maybe it is the workload''; simulation validation closes ``maybe it is the simulator.'' Applied together, these techniques produce evaluations that stand up to scrutiny.

From a practical perspective, we recommend that student researchers adopt the minimal toolkit of \S\ref{sec:toolkit} as a default and treat the checklist of Appendix~\ref{sec:checklist} as a self-review pass before each submission.

\begin{quote}
\emph{A result is convincing only when it is reproducible, consistent across regimes, and free of artifacts introduced by the experimental setup.}
\end{quote}

\appendix

\section{Pre-submission Checklist}
\label{sec:checklist}

A single-page self-review before sending the paper out. Each item closes a specific class of reviewer objection.

\begin{itemize}[leftmargin=*, itemsep=2pt]
    \item[$\Box$] Hypothesis stated explicitly, in falsifiable form (\S\ref{sec:design}).
    \item[$\Box$] At least one strong baseline (not a strawman) (\S\ref{sec:design}).
    \item[$\Box$] $\geq 10$ seeds per configuration; $\geq 30$ if the effect is small (\S\ref{sec:variability}, \S\ref{sec:sample-size}).
    \item[$\Box$] Variability reported via 95\% CIs, preferably bootstrap (\S\ref{sec:bootstrap}).
    \item[$\Box$] Paired non-parametric test used where the design allows (\S\ref{sec:hypothesis-testing}).
    \item[$\Box$] Effect sizes reported alongside $p$-values, with CIs on differences (\S\ref{sec:hypothesis-testing}).
    \item[$\Box$] At least two regimes evaluated; results reported per regime, not averaged (\S\ref{sec:multifactor}).
    \item[$\Box$] Where the method does \emph{not} help is stated explicitly (\S\ref{sec:weak-vs-strong}).
    \item[$\Box$] Multiple-comparison correction applied where many tests are reported (\S\ref{sec:multiple-comparisons}).
    \item[$\Box$] Latency reported with median + tail percentiles + CDF; warmup discarded (\S\ref{sec:latency-pitfalls}).
    \item[$\Box$] Error bars are 95\% CIs, not SE or SD; captions self-contained (\S\ref{sec:reporting}).
    \item[$\Box$] Held-out workloads or conditions for out-of-sample validation (\S\ref{sec:out-of-sample}).
    \item[$\Box$] Simulator passes sanity checks, baseline reproduction, sensitivity, conservation (\S\ref{sec:sim-validation}).
    \item[$\Box$] Common random numbers used to pair runs across policies (\S\ref{sec:sim-validation}).
    \item[$\Box$] Seeds, hardware, code, and workload generators released or described (\S\ref{sec:reproducibility}).
\end{itemize}

\section{Glossary}
\label{sec:glossary}

\begin{description}[style=nextline, font=\bfseries, leftmargin=0pt, itemsep=2pt]
    \item[Bootstrap] Resampling-based estimator of a sampling distribution. Resample the data with replacement many times, recompute the statistic, and use the resulting empirical distribution to form CIs. Makes no parametric assumption.
    \item[Confidence interval (CI)] A range that, under repeated experiments, contains the true parameter with a stated probability (e.g., 95\%). A CI on a difference that excludes zero is the visual analogue of a significant test result.
    \item[Cliff's delta] Non-parametric effect size for two ordinal samples, equal to the probability that a random value from one group exceeds one from the other, minus the reverse. Robust to heavy tails.
    \item[Common random numbers (CRN)] Variance-reduction technique: drive multiple policies with the same random sequence so per-trial differences isolate the policy effect. Enables paired tests.
    \item[Coordinated omission] Measurement artifact in closed-loop benchmarks: when the system stalls, request dispatch is paused, so the worst latencies are never recorded. Latency relative to \emph{intended} dispatch time corrects for this.
    \item[Effect size] Magnitude of a difference, expressed in interpretable units (\% improvement, absolute difference, Cliff's delta). Distinct from statistical significance.
    \item[Factorial design] Experiment varying multiple factors over a grid of levels, allowing estimation of interactions between factors.
    \item[FDR (false discovery rate)] Expected proportion of false positives among reported positives. Controlled by the Benjamini--Hochberg procedure; less conservative than family-wise error rate.
    \item[Forking paths] Inflation of false positives caused by analyst choices (which metric, which subset, which cutoff) made after seeing the data, even without explicit $p$-hacking.
    \item[Power] Probability that a test correctly rejects the null when a true effect of a given size is present. Determined by sample size, effect size, and significance level.
    \item[$p$-value] Probability of observing a result at least as extreme as the data, assuming the null hypothesis is true. A screening criterion, not a measure of effect importance.
    \item[Wilcoxon signed-rank] Non-parametric paired test on the signed ranks of differences. Default replacement for the paired $t$-test when normality is not assured.
\end{description}

\bibliographystyle{plainnat}
\bibliography{stateval}

\end{document}